\documentclass[pre,preprint]{revtex4-1}
\usepackage{amsmath,amsfonts,amssymb}
\usepackage{ucs}
\usepackage[utf8x]{inputenc}
\usepackage[english]{babel}
\usepackage{fontenc}
\usepackage{graphicx}
\usepackage{subfigure}
\usepackage[dvips]{hyperref}

\begin{document}

\title{Exact solutions for a periodic assembly of bubbles in a Hele-Shaw channel}

\author{Giovani L. Vasconcelos$^{1,2}$}
\email{g.vasconcelos@imperial.ac.uk, giovani.vasconcelos@ufpe.br}

\author{Christopher C. Green$^2$}
\email{christopher.c.green05@imperial.ac.uk}

\affiliation{$^1$Department of Mathematics, Imperial College London, 180 Queen's Gate, London SW7 2AZ, United Kingdom}
\affiliation{$^2$Departamento de F\'{\i}sica,  Universidade Federal de Pernambuco, 50670-901, Recife, Brazil}

\begin{abstract}
Exact solutions are reported for a periodic assembly of bubbles steadily co-travelling in a Hele-Shaw channel. The solutions are obtained as conformal mappings from a multiply connected circular domain in an auxiliary complex plane to the flow region in a period cell. The conformal mappings are constructed using the generalized Schwarz-Christoffel formula for multiply connected polygonal domains in terms of products of Schottky-Klein prime functions. It is shown that previous solutions for multiple steady bubbles in a Hele-Shaw cell are all particular cases of the solutions described herein. Examples of specific bubble configurations are discussed.
\end{abstract}

\maketitle

\section{Introduction}

This paper derives analytical solutions for the free boundary problem describing a periodic configuration of an assembly of bubbles in a Hele-Shaw channel---an apparatus in which viscous fluid is confined between two closely-spaced, parallel glass plates. The problem we consider is cast as a potential theory problem, i.e. the flow is governed by Darcy's law and the velocity potential is harmonic. This makes the problem amenable to complex variable methods \cite{howison}. Streams of bubbles moving in such a constricted geometry are of interest in several contexts, such as in oil recovery (e.g. secondary injection at high pressures), in bio-engineering  (e.g. blood oxygenation), and in the related problem of blood flows in narrow vessels \cite{Max, Fu}.
 
The steady motion of bubbles in a Hele-Shaw cell has been very well-studied. Here we focus attention on periodic configurations and neglect surface tension effects on the bubble boundaries. Burgess and Tanveer \cite{BT} determined a family of solutions for an infinite stream of bubbles in a Hele-Shaw channel with one symmetric bubble per period cell. Subsequently, the Burgess-Tanveer solution was generalized to include an arbitrary number of symmetrical bubbles per period cell \cite{GLV1994,prsa2011}. More recently, periodic solutions for a single bubble per period cell with no symmetry constraints have also been obtained \cite{pre2013}. 

The solutions presented in this paper describe a periodic array of bubbles with an arbitrary number of bubbles per period cell with no symmetry assumptions being enforced on the interface shapes; hence they generalize all previous periodic solutions mentioned above. Given that  the flow domain (in the period cell) is of arbitrary connectivity, the problem of determining the multiple bubble interfaces becomes highly non-trivial. Crucial to our method is the choice of a \textit{rectangular} period cell which allows us to make use of a recent result from complex analysis, namely the generalized Schwarz-Christoffel mapping for multiply connected polygonal regions \cite{crowdy1,GLV2014}. 
The present work also offers the first generalization of the recent analytical solutions found by the authors for a finite assembly of bubbles in the Hele-Shaw channel \cite{GV2014} where similar methods were adopted.

\section{Problem formulation}
\label{sec:2}

\begin{figure}[t]
\centering 
\subfigure[\label{fig:1a}]{\includegraphics[width=0.4\textwidth]{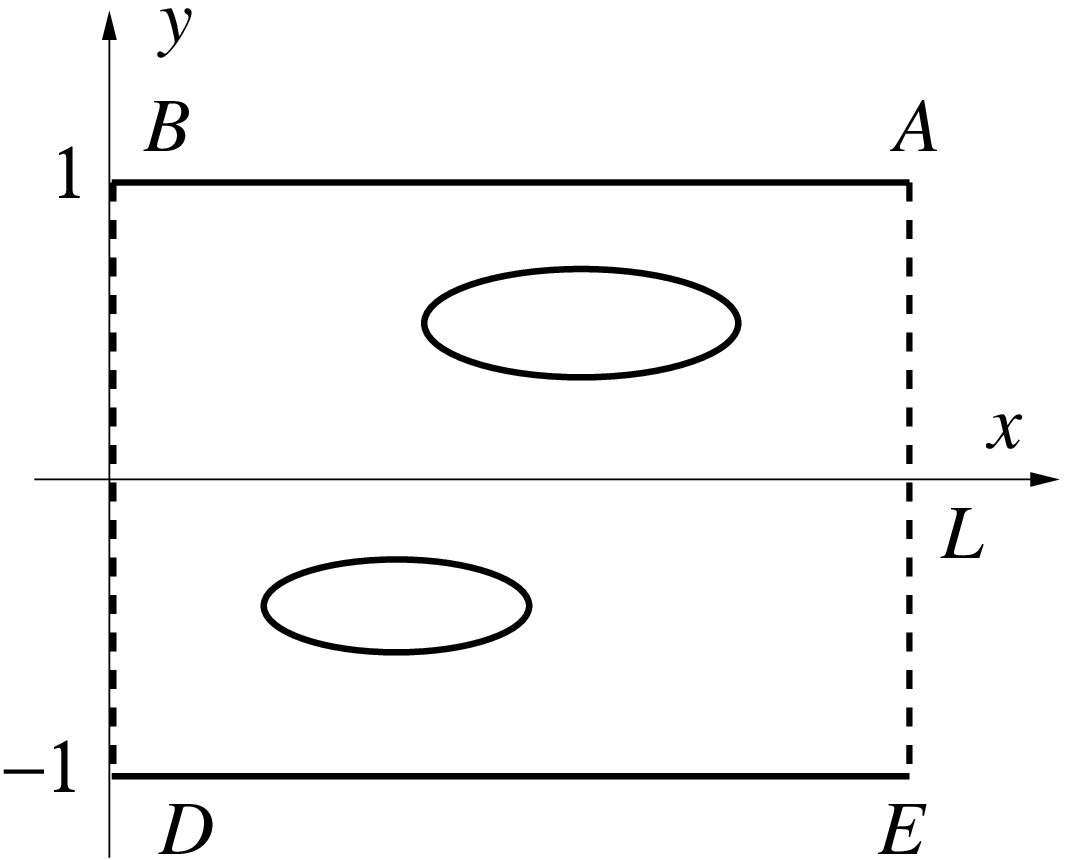}} \quad
\subfigure[\label{fig:1b}]{\includegraphics[width=0.43\textwidth]{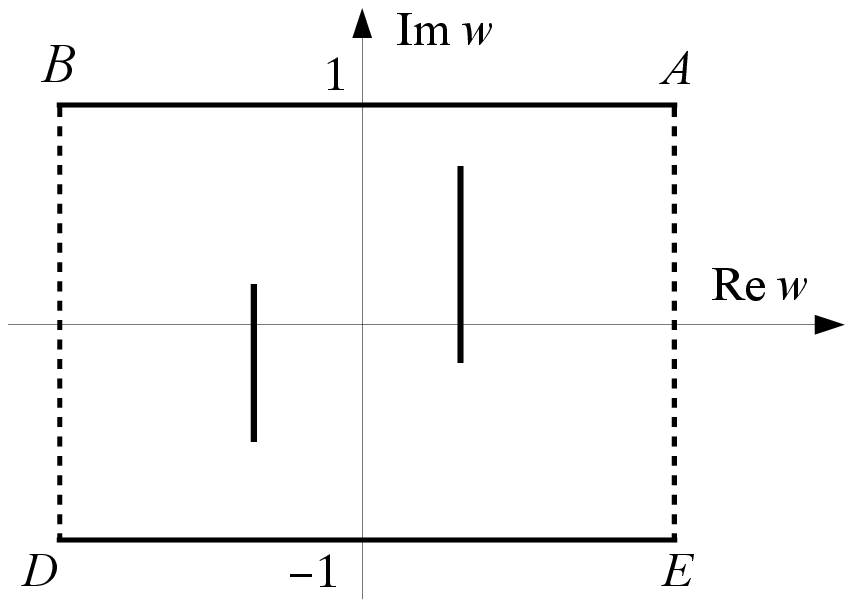}}
\subfigure[\label{fig:1c}]{\includegraphics[width=0.42\textwidth]{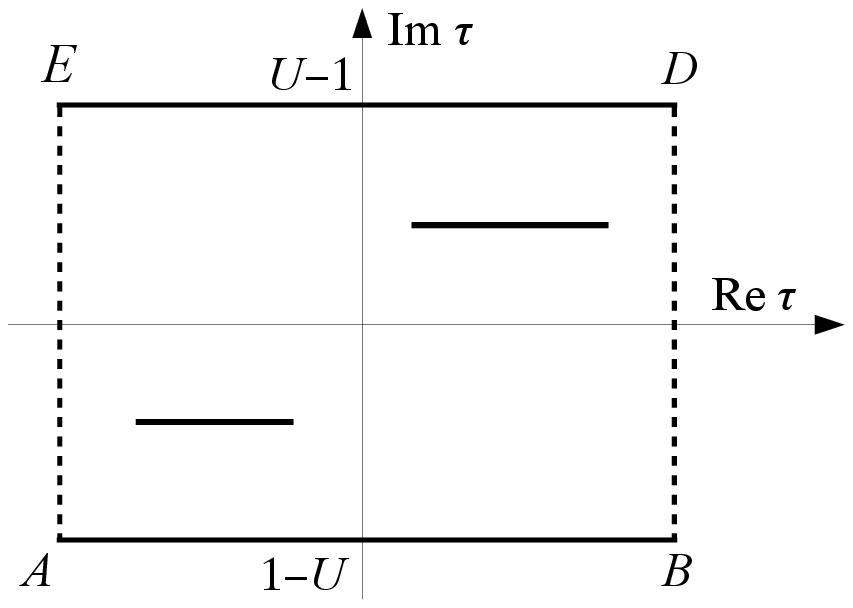}} \qquad
\subfigure[\label{fig:1d}]{\includegraphics[width=0.33\textwidth]{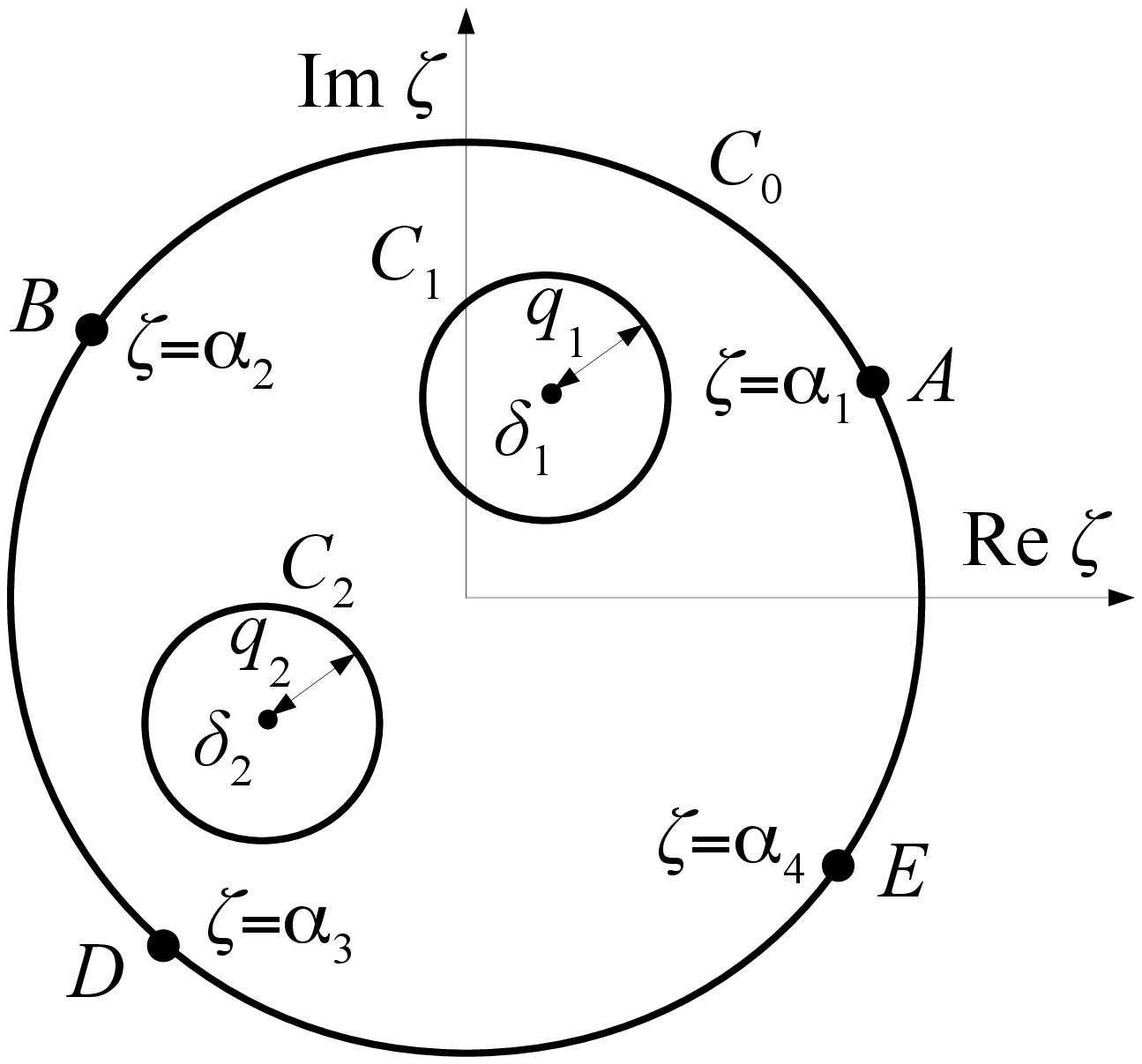}}
\caption{The flow domains for a reduced unit cell: (a) the $z$-plane, (b) the $w$-plane, (c) the $\tau$-plane, and (d) the auxiliary $\zeta$-plane.}\label{fig:1}
\end{figure}

We consider the problem of a periodic assembly of bubbles moving with a constant velocity $U$ in a Hele-Shaw channel whose centerline is chosen to be the $x$-axis. Suppose the channel has width $2$ and the horizontal period of the configuration is $2L$. Owing to the periodicity, it suffices to consider a rectangular period cell with dimensions $2 \times 2L$ whose vertical midline is taken to be the $y$-axis. We suppose furthermore that both the $y$-axis and the period cell edges $y=\pm L$ are equipotentials of the flow. As a result, the flow domain can be further reduced to a unit cell corresponding to one half of the original period cell which we suppose contains an arbitrary number $M$ of bubbles; see Fig.~\ref{fig:1a}. (The full period cell can be obtained by simply reflecting the reduced cell in one of its lateral edges). 

As is well-known, Hele-Shaw flows \cite{howison} are most conveniently described in terms of analytic functions of the complex variable $z=x+\mathrm{i}y$. Introduce the complex potential $w(z)=\phi(x,y)+\mathrm{i}\psi(x,y)$, where $\phi=-p$ is the velocity potential in normalized units, with $p$ being the fluid pressure, and $\psi$ is the associated streamfunction. Let $D_z$ denote the region occupied by the viscous fluid within the unit cell and let $\partial D_j$, $j=1,...,M$, denote the bubble interfaces. The complex potential $w(z)$ must be analytic everywhere in $D_z$ and satisfy the following boundary conditions on the bubble interfaces: 
\begin{align}
{\rm Re}[w(z)]=\mbox{constant}, \qquad z\in \partial D_j, \  \  j=1,...,M.
\label{eq:cc1}
\end{align}
This follows from the fact that the pressure inside the bubbles is constant and that surface tension effects have been neglected. As the channel walls are streamlines of the flow we must have ${\rm Im}[w(z)]=\pm V$ on $y=\pm 1$, where $V$ is the average fluid velocity across the unit cell in the $x$-direction;  without loss of generality we take $V=1$. Furthermore,  ${\rm Re}[w(z)]=\mbox{constant}$ on $x=0,L$, because the period cell edges are equipotential lines. From these boundary conditions, one readily sees that the flow domain in the $w$-plane is a rectangle with $M$ vertical interior slits, each slit corresponding to a bubble; see Fig. \ref{fig:1b}.

Let us now introduce the complex potential $\tau(z)$ in a frame of reference co-travelling with the bubbles:
\begin{align}
\tau(z)=w(z)-Uz.
\label{eq:tw}
\end{align}
As the bubbles are streamlines of the flow in the co-travelling frame, it follows that
\begin{align}
{\rm Im}[\tau(z)]=\mbox{constant}, \qquad z\in \partial D_j, \  \  j=1,...,M.
 \label{eq:cc2}
\end{align}
It is also clear that in the moving frame, the channel walls and the period cell edges remain streamlines and equipotentials of the flow, respectively. One thus concludes that the flow domain in the $ \tau$-plane is a rectangle with $M$ horizontal interior slits, as shown in Fig. \ref{fig:1c}.

Next, consider a conformal mapping $z(\zeta)$ from a bounded multiply connected circular domain $D_\zeta$ in an auxiliary complex $\zeta$-plane  to the flow domain $D_z$ in the $z$-plane. We choose $D_\zeta$ to consist of the unit disc with $M$ smaller discs excised from it; see Fig.~\ref{fig:1d}. Label the unit circle by $C_0$ and the $M$ inner circular boundaries by $C_1,...,C_M$. Denote the centre and radius of the circle $C_j$ by $\delta_j$ and $q_j$, respectively. We take $z(\zeta)$ to map the unit circle onto the outer boundaries of the unit cell $D_z$, while the inner circles $C_j$, $j=1,...,M$, are mapped to the bubble interfaces; see Figs.~\ref{fig:1a} and \ref{fig:1d}.

If we define the following functions
\begin{align}
W(\zeta)\equiv w(z(\zeta)) \qquad \mbox{and} \qquad T(\zeta)\equiv \tau(z(\zeta)),
\end{align}
it then follows from (\ref{eq:tw}) that the conformal mapping $z(\zeta)$ can be written as
\begin{equation}
z(\zeta)=\frac{1}{U}\left[W(\zeta)- T(\zeta)\right]. 
\label{eq:z}
\end{equation}
We have thus reduced our free boundary problem to the much easier task of finding two analytic functions, $W(\zeta)$ and $T(\zeta)$, which map $D_\zeta$ onto rectangular slit domains. This task is carried out in the next section.

\section{The general solution}

The key observation about the functions $W(\zeta)$ and $T(\zeta)$ is that they map the circular domain $D_\zeta$ onto multiply connected degenerate polygonal domains; see Figs.~\ref{fig:1b} and \ref{fig:1c}. This means that $W(\zeta)$ and $T(\zeta)$ can be computed with the help of generalized Schwarz-Christoffel mappings for multiply connected polygonal domains; we will briefly review these before presenting our general solution.

\subsection{Generalized Schwarz-Christoffel mappings}

Consider a bounded $(M+1)$-connected polygonal domain in the $z$-plane. Label the outer boundary of this polygonal domain by $P_0$ and the inner polygonal boundaries by $P_j$, $j=1,...,M$. Let $z(\zeta)$ be a conformal mapping from the circular domain $D_\zeta$ in the $\zeta$-plane onto this polygonal region, such that the unit circle $C_0$ is mapped to the outer polygon $P_0$ and the interior circles $C_1,...,C_M$ are mapped to the inner polygons $P_j$, $j=1,...,M$. Denote by $\{a_{k}^{(j)}\in C_j~|~k=0,1,...,n_j\}$ the preimages in the $\zeta$-plane of the vertices of polygon $P_j$, and let $\pi \beta_k^{(j)}$ be the turning angles \cite{Driscoll} at the respective vertices.

It is shown by Vasconcelos \cite{GLV2014} that the derivative, $z_\zeta(\zeta)$,  of the mapping function $z(\zeta)$ 
is given by the following formula:
\begin{align}
z_{\zeta}(\zeta)
&=\mathcal{B} \,
\frac{ \omega_\zeta(\zeta,1)\omega(\zeta, -1)-\omega_\zeta(\zeta,-1)\omega(\zeta,1)}{\prod_{j=1}^{M}\omega(\zeta,\gamma_{1}^{(j)}) \omega(\zeta,\gamma_{2}^{(j)})}
\prod_{j=0}^{M} \prod_{k=1}^{n_j} \left[\omega(\zeta,a_{k}^{(j)})\right]^{\beta_k^{(j)}}.
\label{eq:SC6}
\end{align}
Here, $\mathcal{B}$ is a complex constant and $\omega(\zeta,\gamma)$ is the Schottky-Klein prime function associated with $D_\zeta$. For a definition of the Schottky-Klein prime function and a discussion of some of its properties, refer to e.g. \cite{crowdy1,GLV2014}. The set of points $\{\gamma_1^{(j)}, \gamma_2^{(j)}\in C_j~|~j=1,...,M\}$ appearing in formula (\ref{eq:SC6}) correspond to the zeros of the following equation \cite{GLV2014}:
\begin{equation}
\omega_\zeta(\zeta,1)\omega(\zeta, -1)-\omega_\zeta(\zeta,-1)\omega(\zeta,1)=0.
\label{eq:gj}
\end{equation}
This equation (\ref{eq:gj}) can be solved numerically once the conformal moduli $\{\delta_j,~q_j~|~j=1,...,M\}$ of $D_\zeta$ have been prescribed.

\subsection{The complex potentials}

As mentioned above, the flow domains in the $w$- and $\tau$-planes are both rectangles with rectilinear slits in their interiors, the only difference being the orientation of the slits---vertical in the former case and horizontal in the latter. Consequently the functional form of the derivatives $W_\zeta(\zeta)$ and $T_\zeta(\zeta)$ is the same: both are special cases of the Schwarz-Christoffel formula given in (\ref{eq:SC6}).

Consider first the case of the function $W_\zeta(\zeta)$. From the rectangular form of the flow domain in the $w$-plane (see Fig.~\ref{fig:1b}), we expect the mapping $W(\zeta)$ to have four square root branch points at four distinct points on $C_0$. Label these points $\{\alpha_k \in C_0~|~k=1,...,4\}$. Each of these points will map to a right-angle vertex in the $w$-plane implying that the corresponding turning angle parameters are $\beta_k^{(0)}=-1/2$, $k=1,...,4$. Furthermore, $W_\zeta(\zeta)$ must have two simple zeros on each of the inner circles $C_j$, $j=1,...,M$, corresponding to the end points of the $M$ interior slits in the $w$-plane. Denote these zeros by $\{a_1^{(j)}, a_2^{(j)}\in C_j~|~j=1,...,M\}$, at which $\beta_1^{(j)}=\beta_2^{(j)}=1$. Using these facts in (\ref{eq:SC6}), one then finds that $W_\zeta$ is given by
\begin{equation}
W_{\zeta}(\zeta)=\mathcal{C} \, \frac{\omega_{\zeta}(\zeta,1)\omega(\zeta,-1)-\omega_{\zeta}(\zeta,-1)\omega(\zeta,1)}{\sqrt{\omega(\zeta,\alpha_1)\omega(\zeta,\alpha_2)\omega(\zeta,\alpha_3)\omega(\zeta,\alpha_4)}} \prod_{j=1}^{M} \frac{\omega(\zeta,a_{1}^{(j)})\omega(\zeta,a_{2}^{(j)})}{\omega(\zeta,\gamma_{1}^{(j)})\omega(\zeta,\gamma_{2}^{(j)})},
\label{eq:W}
\end{equation}
where $\cal C$ is a complex constant.

To ensure that the rectangular cell boundary in the $w$-plane has the correct orientation, we need to enforce the following boundary condition on the unit circle ($\zeta=e^{\mathrm{i}\theta}$):
\begin{equation}
\text{Im}\left[\frac{dW}{d\theta}\right]=\mathrm{Re} \left[\zeta W_\zeta(\zeta) \right]=0, \quad \arg[\alpha_1] < \theta < \arg[\alpha_2].
\label{eq:dW1}
\end{equation}
Similarly, to ensure that the $M$ interior slits are all vertical, we apply the following requirement on the inner circles $C_j$ ($\zeta=\delta_j+q_j \mathrm{e}^{\mathrm{i}\theta}$):
\begin{equation}
\mathrm{Re} \left[\frac{ dW}{d\theta} \right]=\mathrm{Im} \left[(\zeta-\delta_j)W_\zeta(\zeta) \right]=0.
\label{eq:dW2}
\end{equation}
We must also require that $W(\zeta)$ be everywhere single-valued in $D_\zeta$. This implies that a $2\pi$-traversal of an inner circle $C_j$ should return to the same starting point on the $j$-th vertical slit, i.e.,
\begin{equation}
\mathrm{Im}\left[\oint_{C_j}W_{\zeta}(\zeta') d\zeta'\right]=0, \qquad j=1,...,M.
\label{eq:dW3}
\end{equation}

As already mentioned, the derivatives of $W(\zeta)$ and $T(\zeta)$ have the same functional form. Hence
\begin{equation}
T_{\zeta}(\zeta)=\mathcal{K} \ \frac{\omega_{\zeta}(\zeta,1)\omega(\zeta,-1)-\omega_{\zeta}(\zeta,-1)\omega(\zeta,1)}{\sqrt{\omega(\zeta,\alpha_1)\omega(\zeta,\alpha_2)\omega(\zeta,\alpha_3)\omega(\zeta,\alpha_4)}} \prod_{j=1}^{M} \frac{\omega(\zeta,b_{1}^{(j)})\omega(\zeta,b_{2}^{(j)})}{\omega(\zeta,\gamma_{1}^{(j)})\omega(\zeta,\gamma_{2}^{(j)})},
\label{eq:T}
\end{equation}
where $\mathcal{K}$ is a complex constant and the set of points $\{b_{1}^{(j)},b_{2}^{(j)} \in C_j ~|~ j=1,...,M\}$ are the preimages of the end points of the $M$ interior slits in the $\tau$-plane. As before, from the shape of the flow region on the $\tau$-plane, one has the  following boundary condition on $C_0$:
\begin{equation}
\text{Im}\left[\frac{dT}{d\theta}\right]=0, \qquad \arg[\alpha_1] < \theta < \arg[\alpha_2];
\label{eq:dT1}
\end{equation}
whilst on the $M$ inner circles $C_j$, it must hold that
\begin{equation}
\mathrm{Im} \left[\frac{dT}{d\theta} \right]=0.
\label{eq:dT2}
\end{equation}
The function $T(\zeta)$ must of course also be single-valued:
\begin{equation}
\mathrm{Re}\left[\oint_{C_j}T_{\zeta}(\zeta') d\zeta'\right]=0, \qquad j=1,...,M.
\label{eq:dT3}
\end{equation}

\subsection{The conformal map $z(\zeta)$}

In view of (\ref{eq:z}), the desired conformal map $z(\zeta)$ can be written as
\begin{equation}
z(\zeta)=\mathcal{A}+\frac{1}{U}\int_{\zeta_0}^\zeta \left[W_\zeta(\zeta')- T_\zeta(\zeta')\right]d\zeta', 
\label{eq:z2}
\end{equation}
where $\cal A$ is a complex constant of integration and $\zeta_0$ is an arbitrary point inside $D_\zeta$. To obtain a specific solution for $z(\zeta)$, we need to know all the parameters appearing in (\ref{eq:z2}), as discussed next. First, without loss of generality, we can set ${\cal A}=0$ and $\zeta_0$ arbitrarily since this merely fixes the origin. Equally without loss of generality, we can set the velocity of the bubbles to be $U=2$ as solutions for different values of $U$ can be obtained from the $U=2$ solutions by an appropriate re-scaling \cite{prsa2011}. Next, we note that the areas and centroids of the bubbles are mainly governed by the conformal moduli $\{\delta_j, q_j~|~j=1,...,M\}$ of the domain $D_\zeta$ which we take as free parameters. Fixing the conformal moduli allows us to compute the set $\{\gamma_{1}^{(j)},\gamma_{2}^{(j)} \in C_j ~|~ j=1,...,M\}$ by solving (\ref{eq:gj}).

We are then left with $4M+8$ parameters, namely: $\{\alpha_k\in C_0~|~k=1,...,4\}$, $\{a_{1}^{(j)},a_{2}^{(j)}, b_{1}^{(j)},b_{2}^{(j)} \in C_j ~|~ j=1,...,M\}$, and the complex constants $\cal C$ and $\cal K$. By the degrees of freedom afforded by the Riemann-Koebe mapping theorem, we can fix three parameters in the conformal mapping $z(\zeta)$, say, $\alpha_1$, $\alpha_2$, and $\alpha_3$. Furthermore, the value of one extra parameter, say, $\alpha_4$, can be fixed in connection with the period $L$, which we treat as a free parameter. Thus, in constructing specific solutions for the mapping function $z(\zeta)$, we can prescribe {\it a priori} the values of the four $\alpha_k$. The parameters $\{a_{1}^{(j)},a_{2}^{(j)} \in C_j | j=1,...,M\}$ and $\arg[{\cal C}]$ are then obtained by solving (e.g. via a multivariate Newton's method for root finding) the set of $2M+1$ equations given by (\ref{eq:dW1})--(\ref{eq:dW3}). Similarly, the points $\{b_{1}^{(j)},b_{2}^{(j)}\in C_j | j=1,...,M\}$ and $\arg\left(\cal K\right)$ are determined by solving the system (\ref{eq:dT1})--(\ref{eq:dT3}).

Finally, the moduli of $\cal C$ and $\cal K$ are determined {\it a posteriori} by ensuring that the widths of the rectangular cells in the $w$- and $\tau$-planes are both equal to 2:
\begin{align}
\mathrm{Im}\left[\int_{\alpha_3}^{\alpha_2}W_{\zeta}(\zeta') d\zeta'\right]=\mathrm{Im}\left[\int_{\alpha_2}^{\alpha_3}T_{\zeta}(\zeta') d\zeta'\right]=2. 
\label{cond.k}
\end{align}
In the next section, we will  illustrate the foregoing theory by considering some specific examples of bubble configurations. 

\section{Discussion}

\begin{figure}[t]
 \centering
\includegraphics[width=0.45\textwidth]{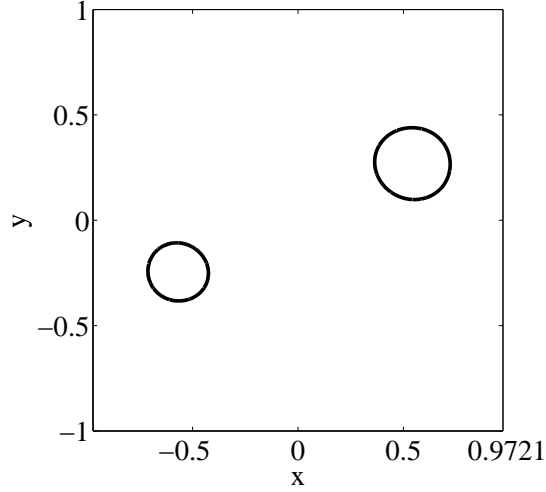}
 \caption{Reduced unit cell with two bubbles. Here the parameters are $\delta_1=-0.525-0.23\mathrm{i}$, $\delta_2=0.5+0.25\mathrm{i}$, $q_1=0.13$ and $q_2=0.16$, with $\alpha_1=\mathrm{e}^{\pi \mathrm{i}/4}$, $\alpha_2=\mathrm{e}^{3\pi \mathrm{i}/4}$, $\alpha_3=\mathrm{e}^{5\pi \mathrm{i}/4}$ and $\alpha_4=\mathrm{e}^{7\pi \mathrm{i}/4}$.}
 \label{fig:2}
 \end{figure}

Solutions for a steady assembly of a finite number of bubbles in a Hele-Shaw channel were recently reported in \cite{GV2014}. These finite-bubble solutions can be obtained as a special case of the periodic solutions presented above by considering the limit $L\to\infty$. This is achieved by setting $\alpha_1=\alpha_4=1$ and $\alpha_2=\alpha_3=-1$, so that these four square root branch points merge pairwise into two logarithmic branch points at $\zeta=\pm 1$ which are respectively mapped to the channel end points \cite{GV2014}. 

Periodic solutions with an arbitrary number of symmetrical bubbles per period cell were found by Silva and Vasconcelos \cite{prsa2011} by reducing the flow domain (on account of symmetry) to a simply connected domain, and then using the standard Schwarz-Christoffel formula. Their solution is readily obtained from our general solution by simply choosing a  symmetric domain $D_\zeta$.  These authors \cite{pre2013} later considered the case of a stream of asymmetric bubbles
with  one bubble per unit cell (i.e. $M=1$), which could  be  handled by the Schwarz-Christoffel mapping for doubly connected domains \cite{Driscoll}. Their asymmetric solution is but a particular case of the general solution presented here. In this case, as the unit cell in the $z$-plane is doubly connected, the circular domain $D_\zeta$ in the auxiliary $\zeta$-plane can be chosen to be a concentric annulus $q<|\zeta|<1$. For this geometry, the Schottky-Klein prime function admits the following simple form \cite{crowdy1}:
\begin{equation}
\omega(\zeta,\alpha)=-\left(\alpha \prod_{n=1}^{\infty}\left(1-q^{2n}\right)^{-2}\right) P\left(\zeta/\alpha,q\right)
\label{omega}
\end{equation}
where
\begin{equation}
P(\zeta,q)=(1-\zeta)\prod_{n=1}^{\infty}\left(1-q^{2n}\zeta\right)\left(1-q^{2n}\zeta^{-1}\right).
\label{pfunc}
\end{equation}
Inserting (\ref{omega}) into (\ref{eq:W}) and (\ref{eq:T}), and using some properties of the function $P(\zeta,q)$, see e.g. \cite{crowdy1}, one can verify that the solution reported in \cite{pre2013} is indeed recovered (albeit in a somewhat different notation).

The formalism we have presented in this paper is very general in that it naturally accounts for any finite number of  bubbles per unit cell with no \textit{a priori} symmetry assumptions concerning the bubble shapes. The physical parameters of the bubble assembly (i.e. the number of bubbles, their areas and centroids) are all encoded in the prescription of the domain $D_\zeta$ over which the Schottky-Klein prime functions appearing in (\ref{eq:z2}) are defined. One example of a specific bubble configuration with two bubbles per unit cell (i.e. $M=2$) is shown in Fig.~\ref{fig:2} for the parameters $\delta_1=-0.525-0.23\mathrm{i}$, $\delta_2=0.5+0.25\mathrm{i}$, $q_1=0.13$ and $q_2=0.16$, with $\alpha_1=\mathrm{e}^{\pi \mathrm{i}/4}$, $\alpha_2=\mathrm{e}^{3\pi \mathrm{i}/4}$, $\alpha_3=\mathrm{e}^{5\pi \mathrm{i}/4}$ and $\alpha_4=\mathrm{e}^{7\pi \mathrm{i}/4}$. Here $L=1.94$. Solutions for a higher number of bubbles (per unit cell) can be obtained in similar manner but the numerical computation of the accessory parameters becomes increasingly more expensive.

\section{Conclusions}

We have presented an exact solution for a periodic assembly of bubbles co-travelling in a Hele-Shaw channel with an arbitrary number of bubbles per period cell. The solution is obtained as a conformal map $z(\zeta)$ from a multiply connected circular domain in an auxiliary complex $\zeta$-plane to the flow domain in the reduced period cell.
The  mapping function is written explicitly in integral form in terms of products of Schottky-Klein prime functions which can be computed with very accurate algorithms \cite{thesis}. It was shown that all previous solutions for multiple steady bubbles in a Hele-Shaw cell can be viewed as particular cases of the solutions described here. An interesting extension of the present work would be to consider a periodic array of bubbles where the period cell is not necessarily rectangular. In this general setting, Schwarz-Christoffel mappings can not be deployed and a new mathematical approach is needed. Work in this direction is currently in progress.

\begin{acknowledgments}
CCG is appreciative of the hospitality of the Department of Physics at the Federal University of Pernambuco where part of this work was carried out. GLV thanks the Department of Mathematics at Imperial College London, where this work was completed, for its hospitality during a sabbatical stay. CCG acknowledges financial support from a Doctoral Prize Fellowship of  EPSRC (United Kingdom). GLV acknowledges financial support from a scholarship of CNPq/CsF (Brazil).
\end{acknowledgments}

\end{document}